\documentclass[doublecol]{epl2} 
 \usepackage{graphics}
 \usepackage{graphicx}
\usepackage{amsmath}%
\usepackage{amssymb}%

\newcommand{\kag}{kagom\'e\ }

\title{Low energy excitations of the \kag antiferromagnet and the spin gap issue.}
\shorttitle{
Kagom\'e antiferromagnet low-T dynamics and spin gap issue.
 } 

\author{P. Sindzingre 
\thanks{ E-mail:
 \email{phsi@lptmc.jussieu.fr}
} 
\and C.~Lhuillier 
\thanks{ E-mail:
 \email{claire.lhuillier@upmc.fr}
} 
}
 \institute{Laboratoire de Physique Th\'eorique de
la Mati\`ere Condens\'ee,\ Univ. P. \& M. Curie, \ CNRS, UMR 7600,\\
Case Courrier 121,\ 4 place Jussieu,\ 75252 Paris Cedex,\ France }
\begin{document}
\title{Low energy excitations of the \kag antiferromagnet and the spin gap issue.}
\shorttitle{
Kagom\'e antiferromagnet, low-T dynamics and the spin gap issue.
 } 

\author{P. Sindzingre 
\thanks{ E-mail:
 \email{phsi@lptmc.jussieu.fr}
}
\and C.~Lhuillier 
\thanks{ E-mail:
 \email{claire.lhuillier@upmc.fr}
}
}
\shortauthor{ P. Sindzingre \and C.~Lhuillier }

%

%

\abstract{
In this paper  we report the latest results of exact diagonalizations 
of $SU(2)$ invariant models on various lattices (square, triangular,
 hexagonal, checkerboard and \kag lattices) and revisit critically  
in this light some of our previous analysis. 
We focus on the low lying levels in each $S$ sector. 
The differences in behavior between gapless systems and gapped 
ones are exhibited.  The plausibility of a gapless spin liquid 
in the Heisenberg model on the \kag lattice is discussed. 
A rough estimate of the spin susceptibility in such an hypothesis is given.
The evolution of the intra-$S$ channel spectra under the effect of a 
small perturbation is consistent with the proximity of a quantum 
critical point.
We reemphasize that the very small intra-$S$ channel 
energy scale observed in exact spectra  is a very interesting  
information to understand the low $T$ dynamics of this model.
}

\maketitle
The RVB concept of a quantum Spin Liquid due to P.W. Anderson in the beginning of the seventies  has motivated a very large number of works both theoretical and experimental~\cite{a73,Misguich2005c}.
 Very recently a specifically interesting candidate for a spin liquid has emerged:
the Herbersmithite characterized by a fluctuating behavior down to temperatures
of the order of  $10^{-4}$ times the coupling
constant~\cite{helton07,mendels2007,bert2007,olariu2008,imai2008}.
  In this compound the magnetic Cu ions are located at the vertices of a \kag
lattice and the interactions between these spin-1/2 ions are supposed to be
fairly isotropic.
The possibility that a simple Heisenberg model might capture the physics
of this compound,  has therefore attracted interest, 
although a significant amount of defects
and other interactions, e.g. Dzyaloshinskii-Moriya interactions,
are likely to be present.
However, in spite of many studies, the nature of the ground-state
and first excitations of the spin-1/2 \kag Heisenberg 
antiferromagnet (KHA) is still 
disputed~\cite{le93,lblps97,web98,ba04,sh2007,ran2007,ryu2007,singh2008,jiang2008}.

 Exact diagonalizations~\cite{ze90,ce92,le93,lblps97,web98} have very early
shown that something special happens in the KHA. n-points correlations
functions  (up to n=8) decrease at short
distances~\cite{e89,ze90,ce92,sh92,le93,nm95,jiang2008}.
The spectral properties observed on these small samples (up to $N=36$ sites)
are quite unusual: these spectra exhibit two low-energy scales:
besides
a finite-size spin-gap scale, 
there is
a second much lower energy scale inside each sector of total spin $S$.
A crude $1/N$ extrapolation of the finite-size spin-gaps to the thermodynamic
limit lead  to  an "estimate" of a putative thermodynamic spin-gap of about 1/20
of the coupling constant~\cite{web98}: an analysis and a result that
we want to revisit critically at the light of recent more extensive studies
of the spectra of samples up to $N=36$ spins.


Many other  numerical approaches  have been used to attack this problem: 
CORE renormalization approaches (Budnik \textit{et al.}~\cite{ba04} and Capponi \textit{et al.}~\cite{capponi2004}),
sophisticated series
expansion (Singh \textit{et al.}~\cite{sh2007,singh2008}),
DMRG calculations (Jiang \textit{et al.}~\cite{jiang2008}),
 and more recently MERA renormalization procedure (Evenbly \textit{et al.}~\cite{Evenbly2009}), give evaluations of  the ground-state energy and for most of them of the spin gap.

 Since 1998 and our publication~\cite{web98} the performances of computers have
increased noticeably (not enough to compute the spectrum of samples much larger
than 36)
but sufficiently for computing a large number of levels in any $S$ sector
of various clusters up to $N=36$ spins
and thus allowing a full analysis of the exact $T=0$
magnetism of this "large molecule".
The object of this paper is to display these results and compare them to
the behaviour of the Heisenberg antiferromagnet on various other lattices:
square (SHA), hexagonal (HHA), triangular (THA), checkerboard (CHA)
of similar sizes.
We show that up to this size, the KHA has a magnetic
behavior which compares extremely well to other antiferromagnets
known to be gapless in the thermodynamic limit and that,
the extrapolation of such finite size data
although it does not rule out a finite spin-gap,
cannot yield a significant measurement of an eventual very small spin gap:
 our 98' paper was on this point misleading\cite{web98}.
From the methodological point of view the analysis of the data that we perform
in this paper can readily be applied to any results obtained by other numerical
finite size calculations (DMRG etc..).
The third message of this paper is that the much lower energy scale that
appears inside each $S$ sector  play a key role in
the true dynamics of this magnet and  suggests a
critical or quasi-critical behavior.

\begin{figure}
\resizebox{0.4\textwidth}{!}{\includegraphics{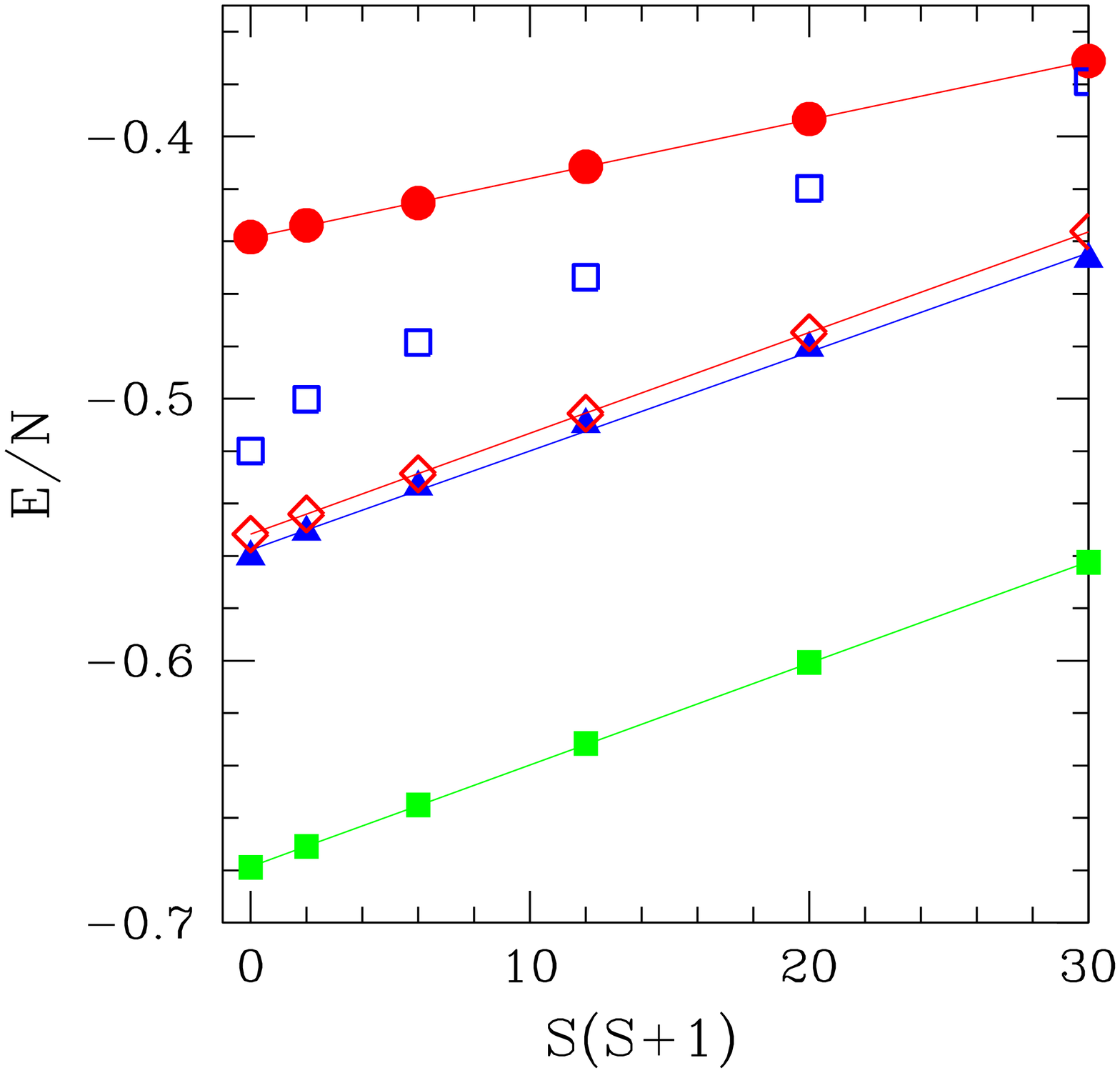}} \\
\resizebox{0.4\textwidth}{!}{\includegraphics{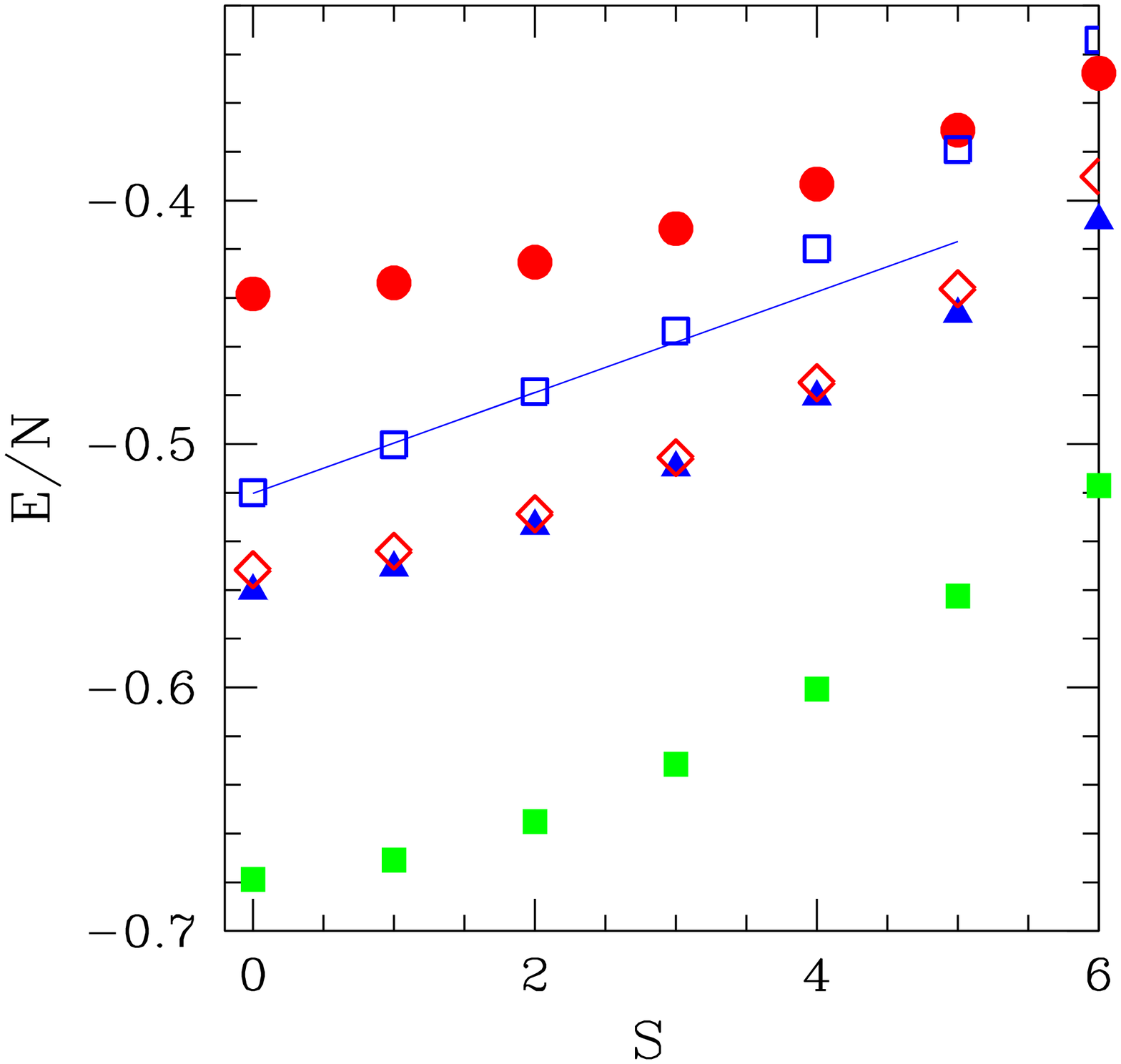}}
\caption[99]{(Color online) Lowest energy per spin of the nearest neighbor Heisenberg model on different lattices versus total spin: square (green squares), triangular (bluetriangles) hexagonal (pink diamonds) 2-dimensional pyrochlore slab(open blue squares) and \kag (red bullets).      Top: exact energies are displayed versus $S(S+1)$. The full lines are linear regressions according to Eq.~\ref{eq:E(S)}. For the pyrochlore lattice  (dot-dashed line) the linear regression only involves the eigenstates with total spins larger or equal to 4.    Bottom: the eigen-levels are displayed versus $S$. The pyrochlore results are linear in $S$ for $S$ ranging from 0 to 3. All samples have 36 sites and the full symmetry of the infinite lattice (periodic boundary conditions).}
\label{fig:1}
\end{figure}

\textit{{Gapless and Gapped magnets through exact diagonalizations:  ground-state energy versus total spin}}

In Fig.~\ref{fig:1}, we display the lowest  eigen-levels  of the
SHA, THA, HHA, KHA and CHA samples of 36 spins  (with periodic boundary conditions) and study how they evolve with the total spin value $S$ of the sample.
The  hamiltonian is
\begin{equation}
\mathcal{H}= \sum_{\langle i,j\rangle} \vec s_i\cdot\vec s_j
\label{eq:H}
\end{equation}
where $\vec s_i$ is the individual spin 1/2 at site $i$ and the  sum ${\langle i,j\rangle}$ runs over pairs of nearest neighbors ($\vec S = \sum_{ i} \vec s_i$).

In the checkerboard lattice, also known as the pyrochlore slab lattice, tetraedra share corners and are disposed on a checkerboard: all couplings along tetraedra edges are identical.

 Fig.~\ref{fig:1} top shows that in most of the cases (except CHA), the evolution of the first level in each $S$ sector is extremely well described by a  $S(S+1)$ behavior which extends to $S=S_{max} =N/2$ (not shown in the figure for the sake of clarity) ~\footnote{In fact there is a tiny cusp at $S=6$ (1/3 of the maximum magnetization)  both for the triangular system and the \kag system which signals a magnetization plateau, and a change of behavior between $S_{max}-1$ and $S_{max}$ which is the manifestation of localized states in the nearly magnetized KHA~\cite{honecker02}. }.

Such a behavior can be described by the equation:
\begin{equation}
 e_N(S)= e_N(0) + \; \alpha_N\;\frac{S(S+1)}{N^2}
 \label{eq:E(S)}
 \end{equation}
 where $e_N(S)$ is the  lowest energy per spin in the $S$ channel for a  sample of size $N$.
The limit of Eq.~\ref{eq:E(S)}, when the size  of the sample goes to infinity is  the well-known  formula of the energy versus magnetization of classical antiferromagnets:
\begin{equation}
 e(m) = e(0) + \;\frac{1}{2\chi}\; m^2
 \label{eq.Landau}
 \end{equation}
 where $m$ is the magnetization ($ 0 \leq m \leq 1$) of the sample and $\chi$ the
 dimensionless susceptibility.  (With this definition, the magnetic moment per spin in a magnetic field $H$ reads  $\mu = (g \mu_B /2)^2 \chi H$ and $\alpha_N =2/\chi_N$).

 The low energy effective behavior described by Eq.~\ref{eq:E(S)} is expected from first principles for a uniaxial  ordered antiferromagnet~\cite{hl90}, as is the case of the SHA and HHA on the square and hexagonal lattices~\cite{fsl01} . In such a case, Eq.~\ref{eq:E(S)}  describes the free precession of the order parameter of the magnet (which has the symmetry of a rigid rotator)  in an $SU(2)$ invariant environment. 
 A slightly more complicated behavior is expected for a biaxial magnet like the THA
(on the triangular lattice): the order parameter has now the symmetry of a quantum top and two different susceptibilities are needed to describe its  free motion~\cite{bllp94}.
It can indeed be observed  in Fig.~\ref{fig:1} that the fit of the low lying levels of the THA with Eq.~\ref{eq:E(S)} is not as good as in the case of collinear N\'eel order (SHA et HHA). In fact the Pearson product-moment correlation coefficient of the fit of the 6 first levels ($S_{tot} =0 \cdots5$) of the exact spectra to Eq.~\ref{eq:E(S)} is 1 for HHA, resp. 0.99997 for SHA, 0.99996 for KHA  and only 0.99849 for  THA.


\begin{figure}
\resizebox{0.4\textwidth}{!}{\includegraphics{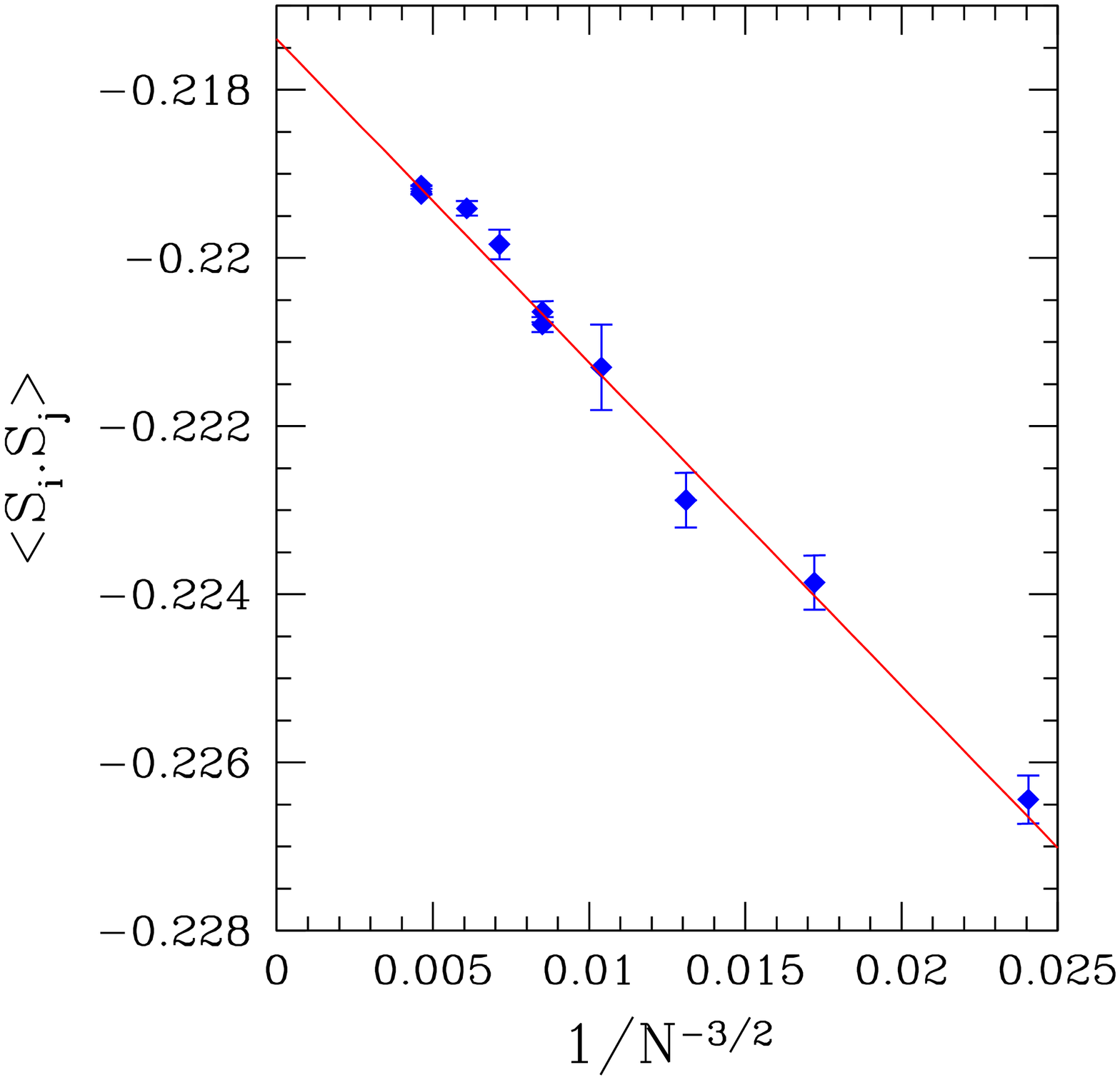}}\\
\resizebox{0.4\textwidth}{!}{\includegraphics{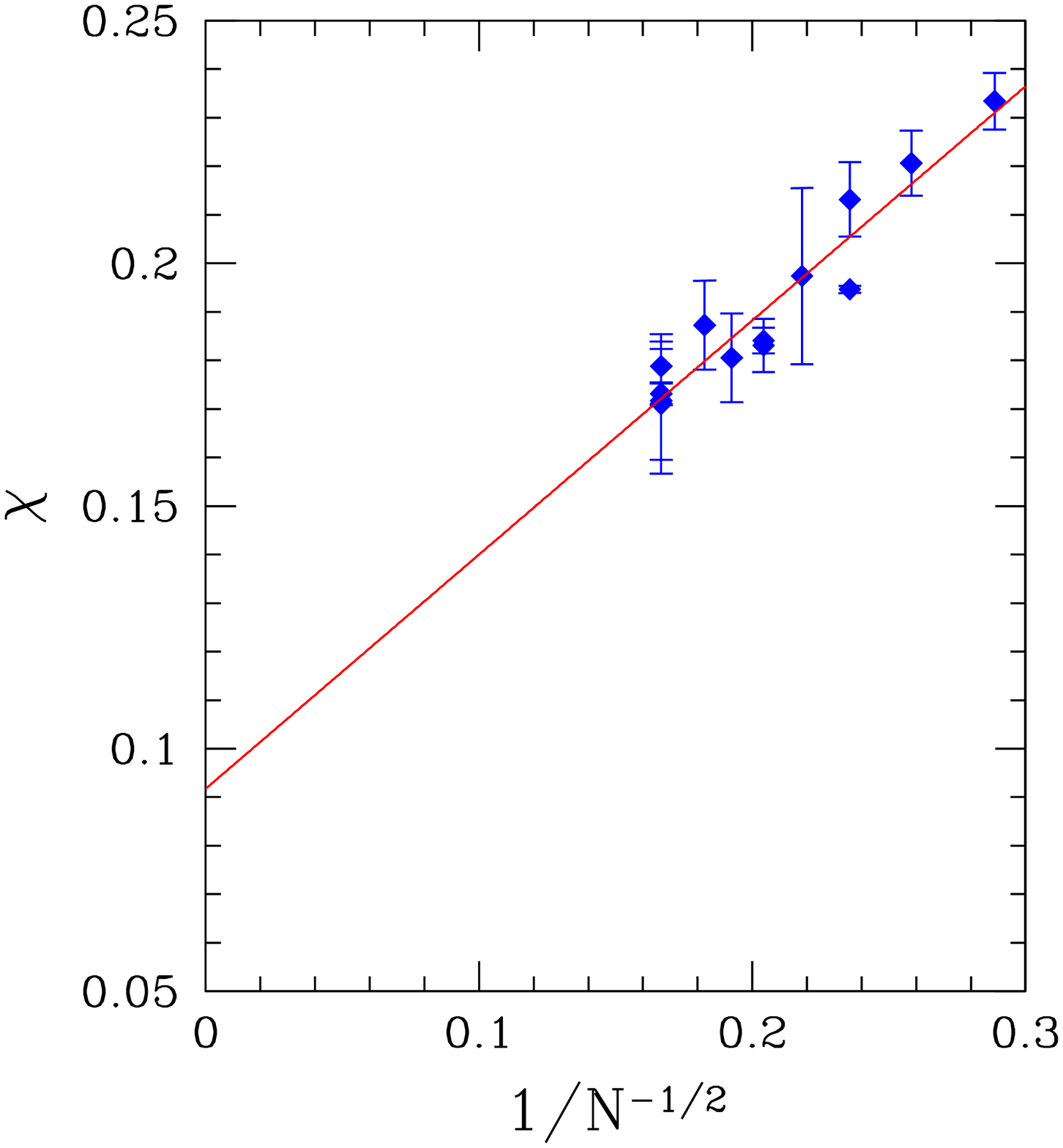}}
\caption{(Color online)
{Top: Evolution of the ground-state energy per bond with the sample size (these valuesare extracted  from the fit of the low lying levels of the different spectra to Eq.~\ref{eq:E(S)}. The variation of the energy with size and shapes of  samples is rather smooth (in spite of the fact that many samples used for this graph  have not all the symmetries of the infinite lattice).  Bottom: evolution of the spin susceptibility with the sample size. The size effects are in first approximation proportional to $1/N^{3/2}$ for the ground-state energy per spin and to $1/N^{1/2}$ for the spin susceptibility. }}
 \label{sizeandshapeseffects}
\end{figure}

  In  addition to the above discussed data, Fig.\ref{fig:1}-top exhibits the low lying levels of the spectrum of the CHA (a two dimensional pyrochlore lattice). This system is a good example of a Valence Bond Crystal with a large gap of the order of 0.7~\cite{c02,Fouet2003}.   The  $S$ dependance of the CHA energy is clearly different from that of the previous models. For high values of the total spin CHA energy displays the standard semi-classical quadratic behavior described in Eq.~\ref{eq:E(S)}. But for $S \leq 4$ the energy of the magnet deviates clearly \textit{downwards} from this quadratic behavior, and  appears  dominated by a linear in $S$ behavior (Fig.\ref{fig:1}-bottom)~\footnote{The downwards deviation is the manifestation that resonances which stabilize the quantum magnets are larger and larger when $S$ decreases, as expected.}.  The following heuristic law is then a good first
   order approximation of the small $S$ behavior:~\footnote{
Indeed, specifically in one dimension, this mean-field approach
and Eq.~\ref{eq.Egapful} is certainly a bit naive.
Non analytic terms dominate the behavior of the
spin  susceptibility just above the spin gap, but the overall thermodynamic
behavior is correct}
 \begin{equation}
 E_N(S) = E_N(0) + \Delta_N \; S.
 \label{eq.Egapful}
 \end{equation}
 where $E_N(S)$ is the total energy of the ground-state of the $N$-sites sample in the $S$ channel.
In a magnetic field this equation becomes
  \begin{equation}
 E_N(S) = E_N(0) + \Delta_N \; S^{z} - g\mu_B H S^{z} +\cdots
 \label{eq.Egapful2}
 \end{equation}
 where $\Delta_N$,  the spin gap ($\sim 0.7$ in the case of the CHA) is a measure of the critical field $H_c$ needed to observe a magnetic answer of the system (in units where $g\mu_B=1$).
At that point  we can do the contact between the usual point of view in term of one-particle excitations and our many body approach. The so-called Bose-Einstein condensation of spin excitations at  $H_c$ translates  in our approach in a collapse of a finite  fraction  of eigen-levels with different total $S^{z}$ when $H=H_c=\Delta$.  In our  36-site CHA sample,
 the linear in $S$  behavior of  Eq.~\ref{eq.Egapful} implies that the eigen-levels with total spin components $S^{z} = 1,\,2,\,3$ (which represent  respectively states with 1, 2, 3 bosons) become degenerate when $H$ reaches $H_c$  .

The  perfect adequation of Eq.~\ref{eq:E(S)} for the  description of the low lying levels of the KHA (for all sizes up to $N=36$)  opens a question:  \textit{is it legitimate to ascribe the finite-size gap of the spectrum of this sample to a true spin gap as it can be perfectly described in an alternate, ultimately gapless picture?} 
Either a saturation of the gap with finite size\footnote{%
In fact neither the Contractor  Renormalization approach~\cite{ba04} (CORE), 
nor the Density Matrix Renormalisation Group~\cite{jiang2008} (DMRG)
do observe a real saturation in the gap with increasing sizes up to 
81 (resp. 108). In that last approach this saturation is only obtained  
for quasi uni-dimensional samples with 3 or 4 legs. 
But the same saturation  has not been exhibited for larger bands. 
The MERA renormalization~\cite{Evenbly2009} approach may be 
promising but has not given up to now an estimate of the gaps.}
or a magnetization curve, as described in Eq.~\ref{eq.Egapful},
should be observed to ascertain a true thermodynamic spin-gap. 
With the knowledge of the exact spectra of the KHA up to $N=36$, 
it seems impossible to decide if KHA is a gapless system or a 
gapped one with an extremely small gap.

  In the hypothesis of a gapless system with linear in $k$ low lying excitations,
  $e(0)$ (resp. $ \chi $) would scale as $N^{-3/2}$ (resp. $N^{-1/2}$).
The fits in Fig.~\ref{sizeandshapeseffects} are not contradictory with
this hypothesis. If this picture is valid up to the thermodynamic limit,
the extrapolated value of the spin susceptibility is
$lim_{N \rightarrow \infty}(\chi_{N}) \sim 0.09 \pm 0.015$.
Within the error bars (which are rather large), the extrapolated value of
the susceptibility compares favorably to the bulk susceptibility measurement
of the Herbertsmithite~\cite{olariu2008}  and  is consistent
with the  high temperature series data~\cite{Misguich2007}.

The large finite-size spin gap appears in this point of view
\textit{as a simple manifestation  of the finite value of the uniform spin susceptibility
and of the spin quantization}. In this aspect the behavior of KHA
is not very different from that of SHA, HHA  or THA. But, as we will show below, the situation
is quite different for the intra-$S$-channel excitations, which do not
suffer of the same limitation due to quantization 
and allow the exploration of a much  lower energy scale.


 \begin{figure}
\resizebox{0.47\textwidth}{!}{\includegraphics{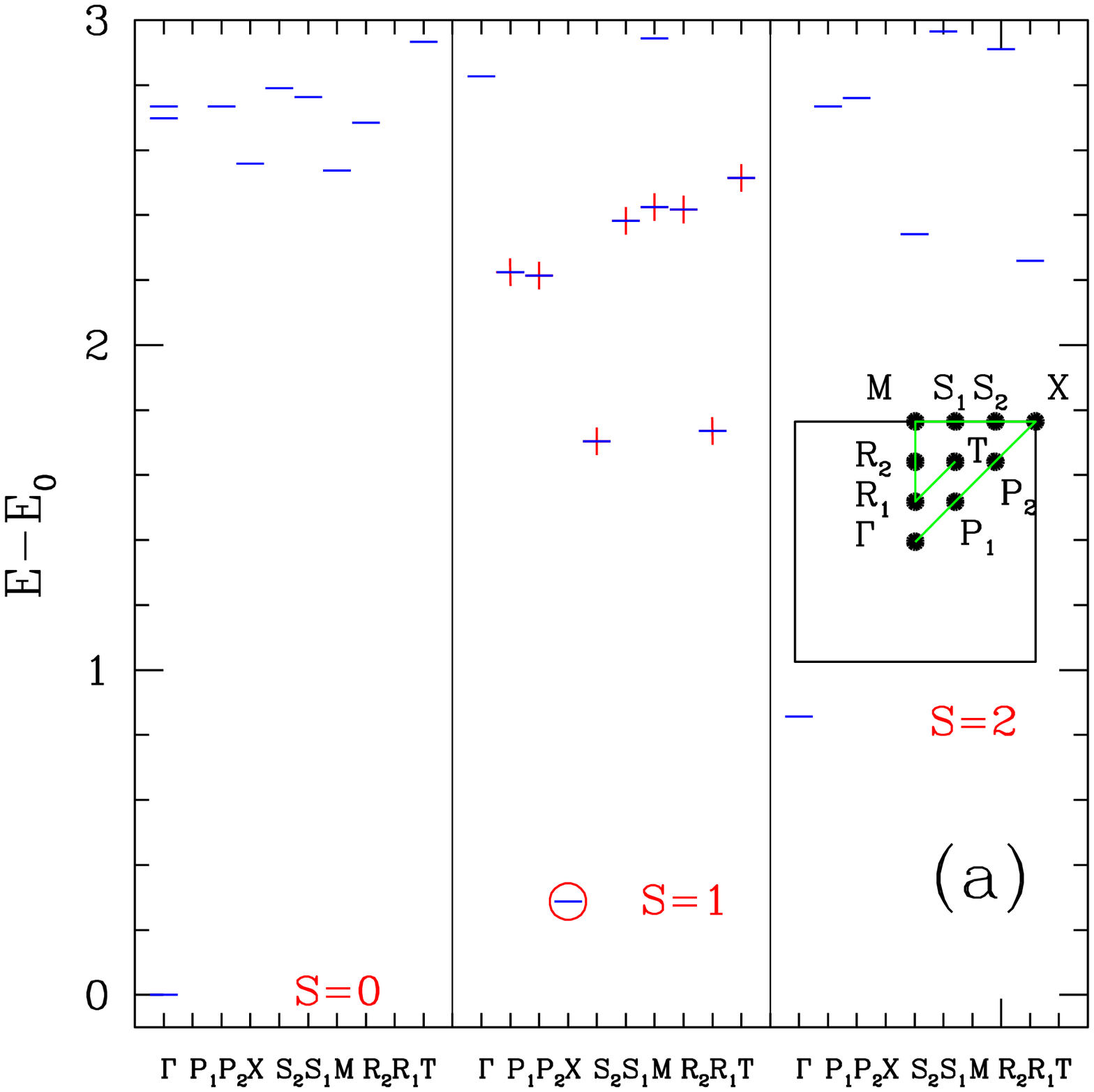}}\\
\resizebox{0.47\textwidth}{!}{\includegraphics{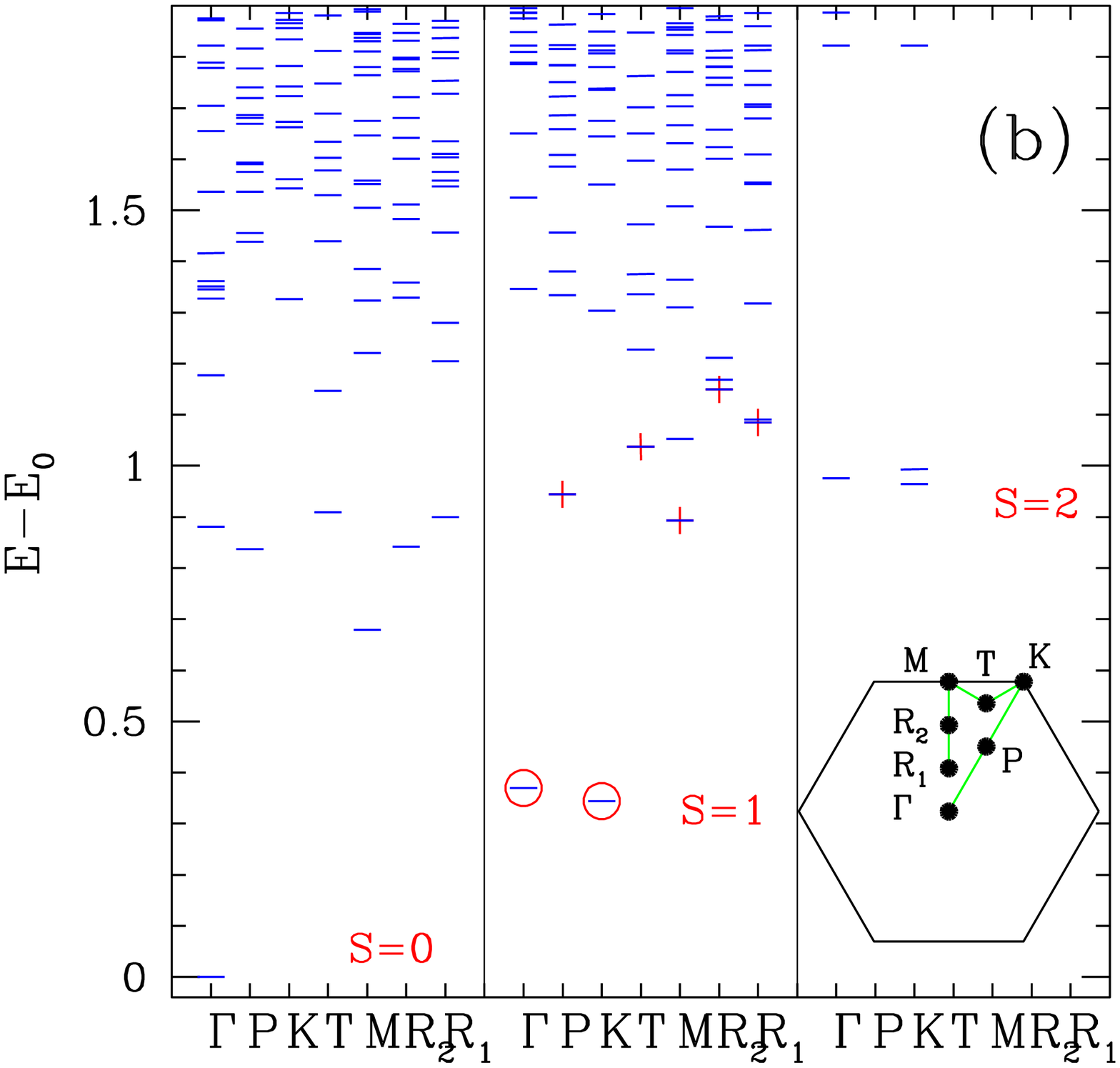}}
\caption{(color online)
Energy above the ground-state of the lowest energy levels,
as a function of wave vectors,
in the lowest spin sectors $S$ 
for the $N=36$ clusters of the square (top) and triangular
lattices (bottom).
  Insets show the location of the wave vectors in the Brillouin zone.
  These clusters have all the symmetries of the infinite lattice.
The full spin wave dispersion  of the square and triangular antiferromagnet
is indicated  by the blue-red crosses; the soft modes by the red circles.
}
 \label{fig:2}
\end{figure}

 \begin{figure}
\resizebox{0.47\textwidth}{!}{\includegraphics{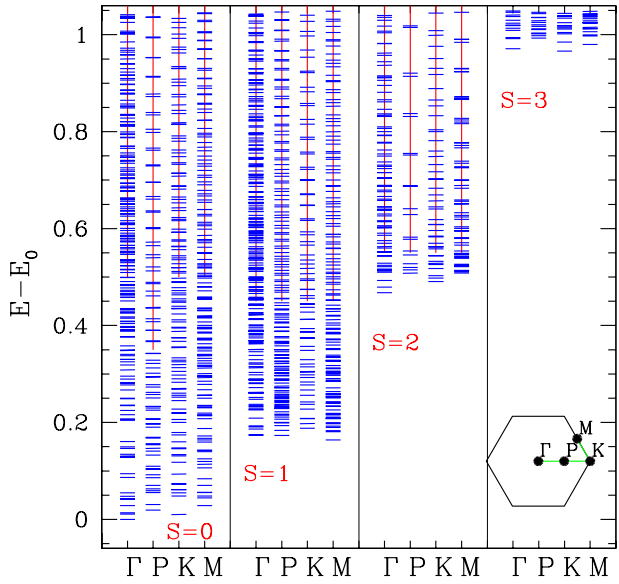}}
  \vspace*{-0.50cm}
\caption{(color online)
Same as~\ref{fig:2} for the \kag lattice.
Note the differences in the vertical scales for these three samples.
The density of low lying states for the kagom\'e sample is a few hundred time
larger than in the two other magnets of the same size.
The convergence of the Lancz\"os process is thus more tedious  and
is only achieved below the vertical red lines.
In the range of non-converged levels, the displayed spectrum is
less dense than in reality.
}
 \label{fig:2c}
\end{figure}

\textit{{N\'eel ordered magnets, energy scale of the excitations}}

   In the  two first  situations SHA and THA (Fig.~\ref{fig:2} a and b), all the low lying levels in each $S$ sector  above the ground-state describe the spin waves.  The analysis of the $S=1$ sector is specially simple~\footnote{The analysis in the $S=0$ sector is more involved, selection rules on the coupling of angular momenta imply that the excited levels in the $S=0$ sectors are uniquely two-magnons excitations when the order parameter is a rigid rotator, whereas for a top- as is the case of the THA- there are both one and two magnons excitations.}: one easily recognizes the soft mode(s) of the SHA  (point X of the first Brillouin Zone (BZ)  Fig.~\ref{fig:2} a) and of the THA (points $\Gamma$ and K of the BZ of the THA Fig.~\ref{fig:2} b).
For the THA the dispersion curve  can directly be compared to the most sophisticated series  calculation~\cite{zheng2006}. The agreement is qualitatively total (flat zone, local minima) and quantitatively rather good.
In spite of the small size of the samples, the order of magnitude of the spin-wave bandwidth  (2 for the SHA and only 1 for the THA) as well as the general behavior of the dispersion curve are in correct agreement with thermodynamic calculations.   Note that in the typical bandwidth of one magnon excitations, the exact spectra in the $S=1$ sector exactly display  one eigen-level per wave number in the SHA case  and approximately the same number in the triangular case (crosses in  Fig.~\ref{fig:2} a and b). (In that latter case the bi-axial nature of the magnet brings some minor complications).

\textit{Specificity of the \kag antiferromagnet}

\begin{figure}
\resizebox{0.4\textwidth}{!}{\includegraphics{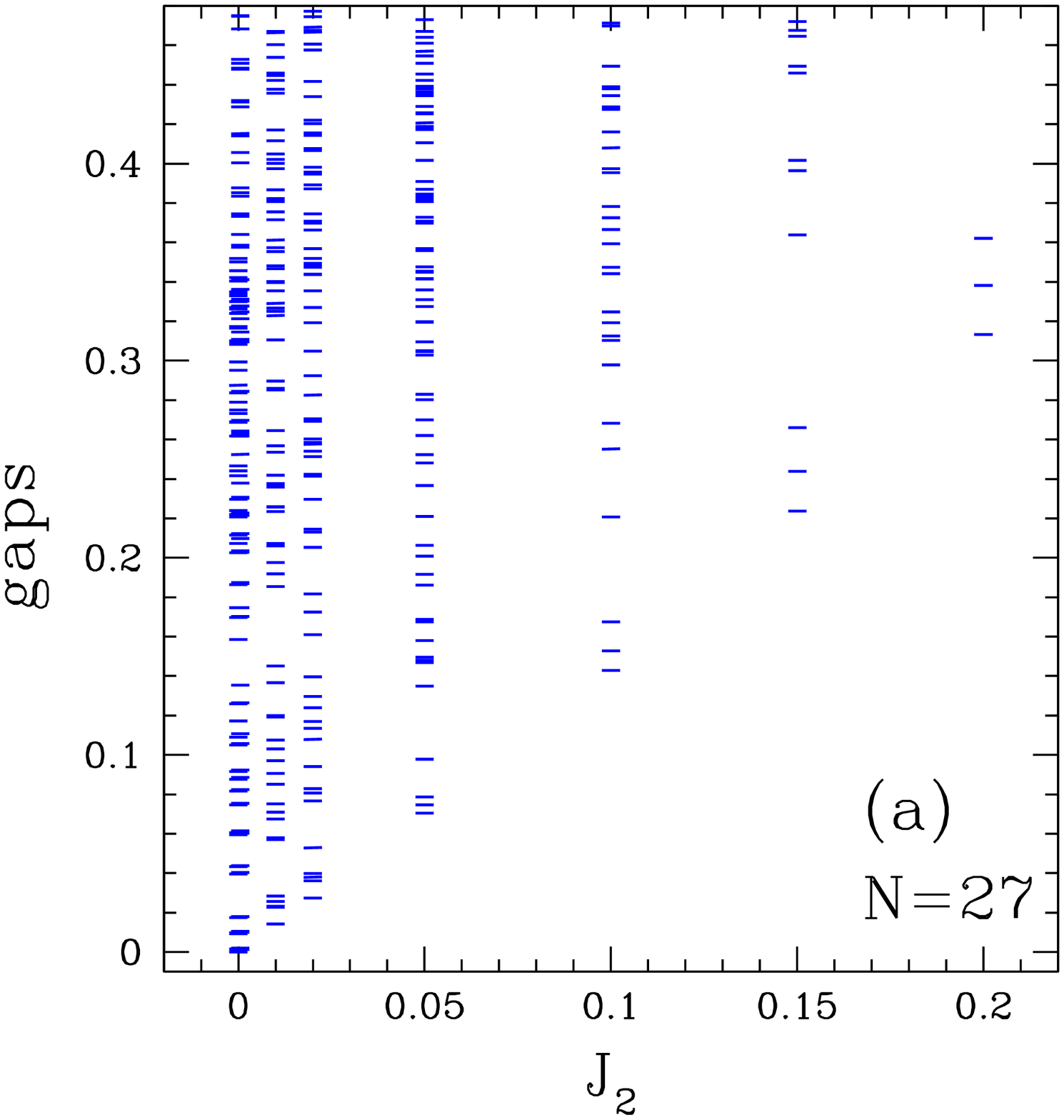}}\\
\resizebox{0.4\textwidth}{!}{\includegraphics{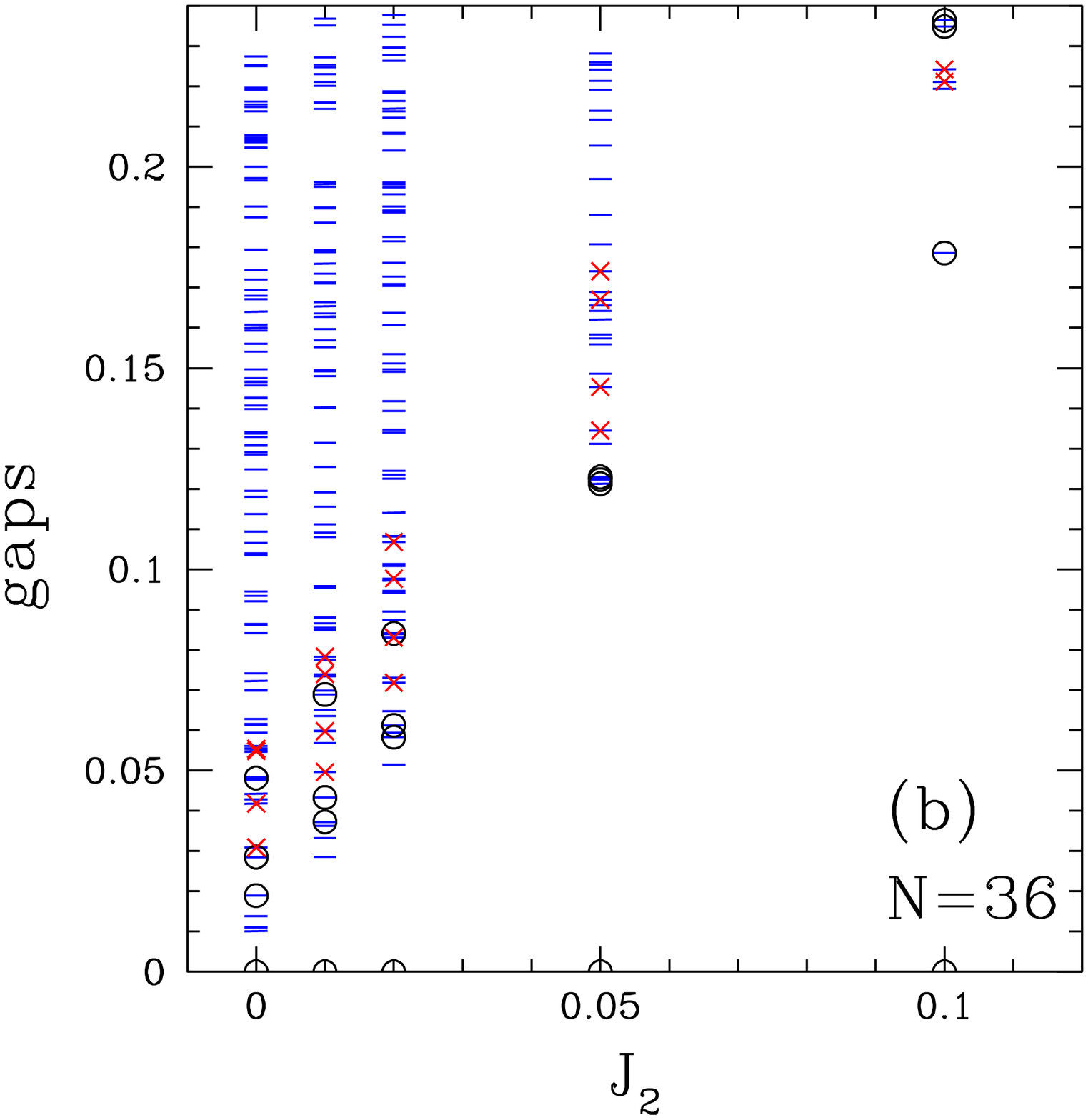}}
\caption{(color online)
Evolution of the low lying levels of the spectrum of the $J_1-J_2$ model
on the kagom\'e lattice versus $J_2$ ($J_1=1$) for a symmetric $N=27$ sample (a)
and for $N=36$ (b).
The total energy of the eigen-levels is measured relative to the absolute 
ground-state for each value of  $J_2$ (black half circles). 
Note that a gap opens quasi linearly just above the ground-state. 
The levels that would embody a VBC-1 order~\cite{sh2007,singh2008,ms07} 
are distinguished by black circles (symmetric pinwheel) 
and red crosses (antisymmetric one).
}
 \label{fig:4}
\end{figure}

    By comparison the spectrum of excitations of the KHA Fig.~\ref{fig:2c}
is qualitatively and quantitatively totally different from the two previous
systems.
It exhibits a considerable density of low lying excitations, seemingly continuous
above the ground-state \textit{in each $S$ sector}.
At same sizes, the finite size gaps  inside an $S$ sector are much smaller.
  A molecule of 36 spins with Heisenberg nearest neighbor couplings
and the coordination symmetry of  a square would have internal Bohr frequencies
for $\Delta S =0$ transitions in the $S=1$ sector of the order of $1.5$.
This number goes down to $0.5$ for the triangular coordination
and to $3.~10^{-3}$ for the \kag network.
This frequency scale at this size is only slightly larger than the one observed in
Herbertsmithite
-in which a fully developed dynamics down to temperatures of the order
of $2.5\;\; 10^{-4}$~\cite{mendels2007,imai2008} has been observed.
The similar low energy of the $\Delta S =0$ transitions in the $S=0$ sector
equally indicate fluctuation of singlets down to very low temperatures.
A measurement of the  energy scale  \textit{inside} each $S$ sector
on larger samples would be of the utmost interest as it might match the experimental scale observed in
Herbertsmithite for $N \sim 108$. Presently impossible with ED it should be reachable in DMRG~\cite{jiang2008}  or MERA~\cite{Evenbly2009}. With some caveat, analysis in the dimer basis could also be interesting~\cite{m98,mm01}.

\textit{Proximity to a quantum critical point?}

The huge density of low lying states in each $S$ sector, and its evolution with the system size~\cite{web98} which is the specificity of KHA, can be interpreted as the absence of an intrinsic energy scale (at least in these small samples). This opens the question of the criticality~\cite{ran2007,ryu2007,hermele2008}.
 We show in Fig.~\ref{fig:4} the evolution of the low lying levels in the $S=0$ (resp.~$S=1/2$) sector of the $N=36$ (resp. $N=27$) sample under the action of an anti-ferromagnetic second
neighbor coupling. Under the action of this perturbation, a gap opens linearly just above the ground-state. The levels that would embody a VBC symmetry breaking~\cite{sh2007,singh2008,ms07} are immediately destabilized as well as all other excited states. The ground-state (resp. in the $S=1/2$ and $S=0$ channels) has all the symmetries expected for the g.-s. of the ${\bf q}=0$  N\'eel order. \textit{At this scale and inside each $S$ sector KAH looks like as it was critical}\footnote{The same situation was observed when destroying the three sublattice long range order on a triangular lattice by a cyclic 4-spin exchange~\cite{LiMing2000}.}.
In refs.~\cite{lblps97,cepas2008}, we concluded that a non zero  critical coupling 
(either a second neighbor interaction~\cite{lblps97}, or
Dzyaloshinski-Moriya~\cite{cepas2008}) was needed to achieve 
full ${\bf q}=0$  N\'eel order.
 These  two determinations are implicitly dependant of the large energy 
scale set by the finite size gap ($\Delta_{N=36}= 0.2$). At the light of the present study on can at best ascertain that the values obtained in these papers are certainly upper-bounds. The present study cannot either ascertain that the system is critical. In some sense it may be instructive to compare the finite-size situation to a non zero temperature regime: in the vicinity of a $T=0$ critical point, it is widely accepted that there exists a $T \neq 0$  quantum critical regime which extents on a finite range of couplings around the quantum critical point. It seems quite reasonable to argue that the behavior observed in refs~\cite{lblps97,cepas2008} around the pure  Heisenberg kagom\'e point looks like a quantum critical regime.  Such an observation does not preclude at lower temperatures/larger sizes  a subsequent crystallization of the pure KAH in a very large Valence Bond Crystal or a small gap RVB phase, but it gives some support to the idea that the system could be near a critical point, and that in a  non negligible range of energy, time and length scale it could behave as "quasi-critical". 
Such a point of view is also supported by the conclusion of  a finite temperature DMFT calculation done some years ago~\cite{gsf01}.

       In this paper we have explicitly shown that the KHA spin-gap measured
in numerics for sizes up to $36$, could be an extrinsic property: 
on such small samples it is impossible to distinguish between a gapless 
system and a system with a very small gap.  
We have shown that the intra-$S$ channel spectra could signal the proximity of a quantum 
critical point. We suggest that polarized inelastic neutron scattering on 
Herbertsmithite  at temperature  low enough but larger than the residual
 Dzyaloshinski-Moriya
perturbation would  bring an interesting insight on the spin-1/2 compound, with a possible comparison to $SrCrGaO$~\cite{mmpfa00}.

\acknowledgments
We are grateful to  IDRIS for the allocation of computer
resources.
We thanks KITP for hospitality during the workshop:
"Moments and Multiplets in Mott Materials"
in 2007  where part of this work was done
and Michael Hermele for stimulating discussions.
 

\end{document}